\documentstyle[useAMS,psfig,referee]{mn2e}
\input{psfig.sty}

\newif\ifAMStwofonts
\def\Mesz{M\'esz\'aros~}
\def\Pacz{Paczy\'nski~}
\title[Cocoons surrounding light, collapsar jets]  
      {Events in the life of a cocoon surrounding a light, collapsar jet}   
\author[E. Ramirez-Ruiz, A. Celotti \& M. J. Rees]
       {Enrico Ramirez-Ruiz$^{1}$, Annalisa Celotti$^{1,2}$ and Martin
      J. Rees$^{1}$ 
\\${\bf 1.}$ Institute of Astronomy, Madingley Road, Cambridge, CB3 0HA.
\\${\bf 2.}$ SISSA, Via Beirut 2-4, I-34014 Trieste, Italy.}

\begin{document}

\maketitle

\label{firstpage}

\begin{abstract}
According to the collapsar model, $\gamma$-ray bursts  are thought to
be produced in shocks that occur after the relativistic jet has broken
free from the stellar envelope. If the mass density of the collimated
outflow  is less than that of the stellar envelope, the jet will then be  
surrounded by a cocoon of relativistic plasma. 
This  material would itself be
able to escape along the direction of least resistance,
which is likely to be the rotation axis of the stellar progenitor, and
accelerate in approximately the same way as an impulsive
fireball.  We
discuss how the properties of  the stellar
envelope have a decisive effect on the
appearance of a cocoon propagating through it. The relativistic
material that accumulated in the cocoon would have enough kinetic energy to 
substantially alter the structure of the relativistic outflow, if
not in fact provide much of the observed explosive power.
Shock waves within this plasma 
can produce $\gamma$-ray and X-ray
transients, in addition to the standard afterglow emission
that would arise from the deceleration shock of the cocoon fireball.
\end{abstract}


\section{Introduction}

Collimated flows of plasma with
velocities close to the speed of light, commonly referred to as
relativistic jets, have been discovered in a
number of astronomical systems. Objects known or suspected to
produce them include: extragalactic radio sources (Begelman, Blandford \&
Rees 1984); microquasars (Mirabel \& Rodriguez 1999);
supernovae (Khokhlov et al. 1999); and $\gamma$-ray bursts
(GRBs). While extragalactic  radio sources produce by far the largest and
most energetic jets in the Universe, GRBs
provide perhaps the most extreme example of relativistic flow that may
be collimated, exhibiting speeds of $\Gamma \approx 100$ or more
(e.g. Waxman, Frail \& Kulkarni 1998; Wang, Dai \& Lu 2000; Panaitescu
\& Kumar 2002; Soderberg \& Ramirez-Ruiz 2002).\\ 
Given the twin requirements of enormous energy $\approx 10^{53}$
erg and association with star forming regions (see M\'esz\'aros 2001 for a
recent review), the currently favoured models all involve massive, collapsing
stars and their byproducts, especially black holes. A ``collapsar''
forms when the evolved core of a massive star collapses to a black
hole, either by fallback or because the iron
core fails to produce an outgoing shock (Woosley 1993; MacFadyen \&
Woosley 1999). The shocks responsible for producing 
the  $\gamma$-rays must arise after the relativistic jet has broken
free from the stellar progenitor, whose density is  reduced along the
rotation axis due to an early phase of accretion\footnote{The majority
of stellar progenitors, with the exception of some very compact stars,
will not collapse entirely during the typical duration of a GRB. A
stellar envelope will thus remain to impede the advance of the
jet (see Matzner 2002).}   
(MacFadyen \& Woosley 1999; Aloy et al. 2000; Wheeler et
al. 2000; MacFadyen, Woosley \& Heger 2001; Wheeler, Meier \& Wilson 2002).  
While the light, relativistic  jet (i.e. light
compared to the stellar density)  makes 
its way out of the progenitor star, its rate of advance is slowed down
and  most of the energy  output during that period is deposited into a
cocoon or ``wastebasket''  surrounding it.  The jet head propagates at mildly
relativistic velocity until it emerges from the edge of the He core
into the low density H envelope at
$r_* \approx 10^{11}$ cm (MacFadyen \& Woosley 1999; \Mesz \& Rees
2001; Matzner 2002). At this stage (and provided that the density of the
H envelope varies steeply with radius; see \S 3), the head of the jet
will advance relativistically, and the cocoon plasma could escape swiftly
from the  stellar cavity and accelerate in approximately the
same way as an impulsive fireball -- its energy will be converted via
adiabatic expansion into bulk kinetic energy.  

Here we describe the evolution and collimation of
such cocoon fireballs. We argue that an understanding of the 
structure and time-dependence of the  cocoon plasma can come only
through a knowledge of the properties of the stellar material through
which it propagates.  A cocoon
fireball may be stalled while propagating through an
extended or a high mass envelope, but  could  expand freely   
beyond the stellar cavity of a Helium post-Wolf-Rayet star.
We show that even if only a small fraction of
the energy in the jet ($\approx$ 1\%) is deposited into the cocoon, it
would have enough kinetic energy to substantially alter the structure of the
expanding outflow, if not in fact provide much of the observed
power. The latter may be true for an observer that lies off-axis
to the jet or at early times when the initial jet contribution to the
emission is negligible. We examine the physical conditions within the
cocoon plasma  -- namely its confinement, collimation and initial entropy -- 
to determine whether or not this delayed fireball will become matter
dominated before it becomes optically thin. We discuss the possible
role of cocoon fireballs in  producing 
$\gamma$-ray and X-ray 
transients (both thermal and non-thermal), along with the standard
afterglow emission that would originate from the deceleration shock. We suggest
that detailed observations 
of this prompt burst and afterglow emission may provide a potential tool
for diagnosing the size of the cocoon cavity and the initial energy to
mass ratio $\eta =(E/Mc^2)$. It also provides a means for probing the
state of the stellar medium through which both the initial jet and cocoon
propagate.

\section{Cocoons: ``Wastebaskets'' of Relativistic Plasma}
 
The properties of the stellar envelope have a decisive effect on
the appearance of a jet propagating through it. The characteristic
stellar progenitor structure is that of an evolved massive star, with
$\approx 2 M_\odot$ Fe core of radius $\approx 10^{9}$ cm and a
$\approx 8 M_\odot$ He core extending out to $r_* \approx
10^{11}$ cm (MacFadyen et al. 2001). In some cases, a cool H envelope
reaches out to $\ge 10^{13}$ cm, while in others the envelope has been largely
lost. The post-collapse (radiation-dominated) pressure profile out to the
edge of the He core drops roughly as $p_{He} \propto \rho_{He}^{4/3} \propto
r^{-2}$ over some two decades in radius (\Mesz \& Rees
2001). Beyond the He core, it drops  drastically as $p_H \propto
r^{\approx -4}$ since at these distances the pressure profile
is still the  precollapse one. The precollapse density of the
presupernova model A25 of MacFadyen et al. (2001) scales roughly as
$\rho \propto r^{-3}$, so that the radiation-dominated pressure is $p
\propto r^{-4}$ (see \Mesz \& Rees 2001). \\
 
\begin{figure*}
\psfig{figure=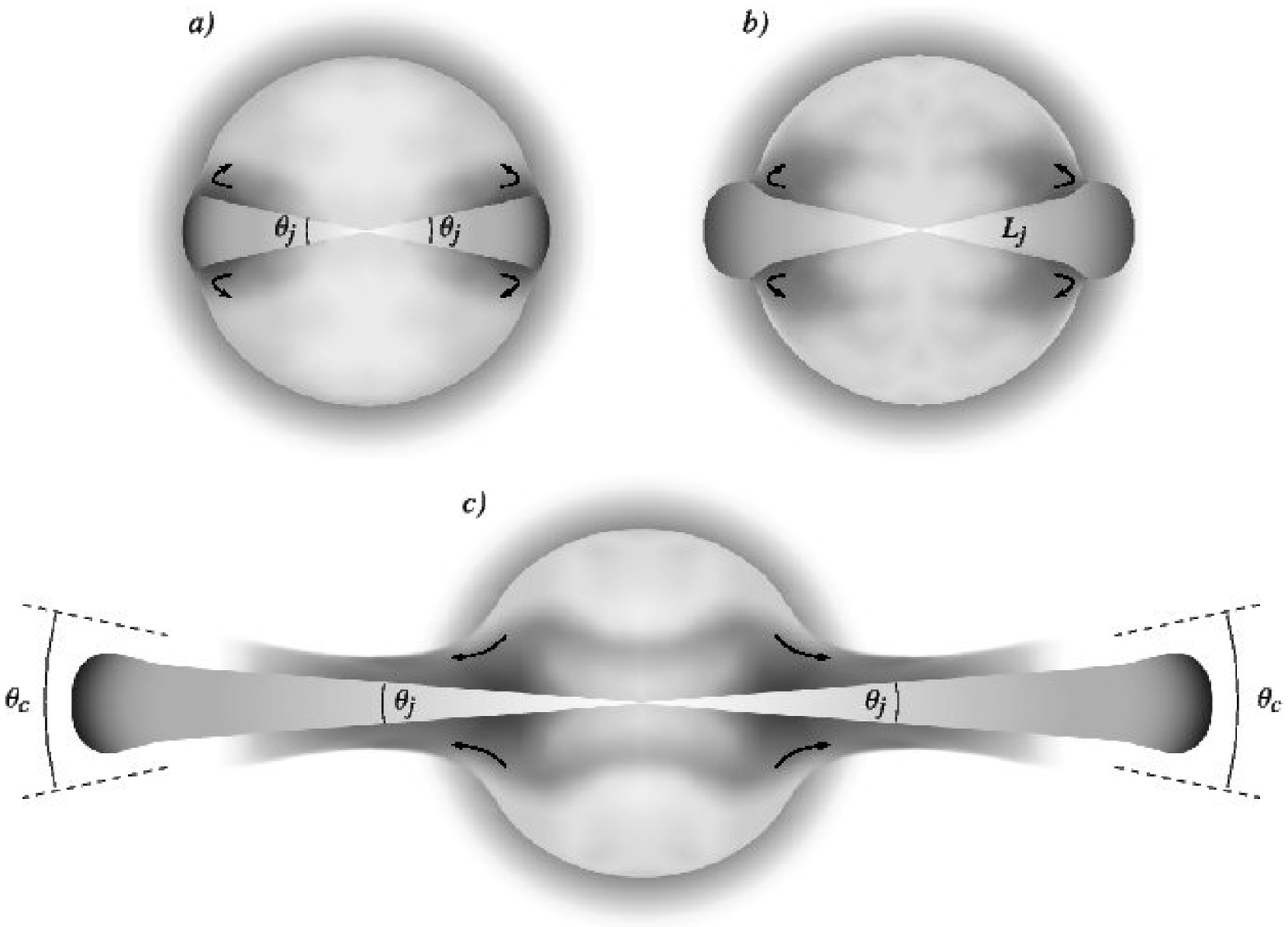,angle=0,width=0.9\textwidth}
{\caption{Schematic diagram illustrating the propagation of a
relativistic jet through the stellar envelope. Initially, the jet is
unable to move the envelope material to a speed comparable to its own
and thus is abruptly decelerated (a). Most of the energy 
output during that period is deposited into a cocoon or
``wastebasket'' surrounding the jet (b). After the jet head advances
relativistically, the cocoon plasma would itself be able to escape swiftly
from the  stellar cavity and accelerate in approximately the
same way as an impulsive fireball (c).}
\label{fig1}}
\end{figure*}

Suppose that a collimated beam has been
established. If all the particles in the beam are ultra relativistic, then
$p_j={1 \over 3}\rho_j c^2$ and the sound speed is $c_s \approx
c/\sqrt{3}$. At a given time, the beam will have evacuated a channel
out to some location  where it impinges on the stellar
envelope\footnote{Reconversion into random energy occurs at the end of
the  channel, which is a natural site for particle acceleration
(Colgate 1974; Chevalier 1982; \Mesz
\& Waxman 2001; Ramirez-Ruiz, MacFadyen \& Lazzati 2002).} at a
``working surface'' which itself advances out 
at speed $V_h$ (see Fig.~\ref{fig1}a). If the power in the jet, $L_j$, is
roughly conserved and stationary, then approximating the channel
as a cylinder of radius $r$, we balance momentum fluxes at the
``working surface'' to obtain  
\begin{equation}
{L_j \over \theta_j^2 r^2 c} \approx \rho_{\rm env}V_h^2
\;\;\;\;\;v \goa V_h \goa c_s,
\end{equation} 
where $v$ is the speed of the beam, $\rho_{\rm env}$, the stellar
envelope density, and $\theta_j$ the jet opening angle. If the beam consists of
relativistic plasma then $v \approx c$, and relativistic
fluid mechanics must be used (the reader is referred for further
details to the generalised formalism developed by Matzner 2002). 
During propagation 
in the iron and He cores, the head of the jet will propagate with
subrelativistic velocities. The energy supplied by the jet exceeds that
imparted to the swept-up stellar material by a factor $\sim
c/V_h \gg 1$. The surplus (or waste) energy must then not 
accumulate near the ``working surface'' but be deposited within a
cocoon surrounding the jet (Fig.~\ref{fig1}; slightly resembling the
cocoons that envelop jets of radio sources: see  
Begelman et al. 1984 and Begelman \& Cioffi 1992). 

After the jet emerges into the H envelope, the
sudden and drastic density drop at the outer edges
permits the jet head  to accelerate to velocities
close to the speed of light ($V_h  
\approx c$). Thus, if it is a general property that the jet
becomes relativistic near the boundary of the He core, the
outer edge of the H envelope is reached in a crossing time $\approx
r_H/2c\Gamma_h^2$ as measured by an observer along the line of sight. 
For an H envelope density varying as $\rho_H \propto
r^{-\beta}$ with $\beta \ge 3$ (see \S 3), the outer edge of the star is
reached in a crossing time that may matter little when compare to the He-core
traversal time\footnote{Many presupernova stars, on the other hand,
have density profiles with $\beta \sim 2$ (Chevalier 1989). For these
stellar progenitors, the jet may be unable to punch through the
stellar envelope (Matzner 2002).}. The fraction of  
relativistic plasma injected  into the cocoon after $V_h \approx c$
will be much  reduced (\Mesz \& Rees 2001). The amount  
of energy that accumulated in the cocoon while the jet was advancing
subrelativistic is then\footnote{A more rigorous estimate of the cocoon
energy content, which accounts for both relativistic and
non-relativistic jet propagation, can be found in equation 10 of
Matzner (2002).}      
\begin{equation}
E_c = \int_0^{t_{\rm He}} L_j(t) dt \approx {r_*  L_j \over \bar{V}_{\rm
h}} \approx 5 \times 10^{50} \bar{V}_{{\rm h},10}^{-1} \;r_{*,11} L_{j,50}
\;\;{\rm erg}, 
\end{equation}
where $t_{\rm He} \approx r_*/\bar{V_{\rm h}}$ is the  He-core traversal
time, $\bar{V_{\rm h}}$, the average speed of the jet head (which is
about c/2; Aloy et al. 2000), and we adopt the convention $Q =
10^x\,Q_x$, using cgs  units. The cocoon expands in the
transverse direction with a velocity given by $V_c \approx
(p_c/\rho_{\rm env})^{1/2}$. The forces driving the cocoon expansion will be
effective so long as  $p_c > p_{\rm env}$ (Matzner 2002). 
The cocoon material could in principle reach pressure equilibrium with
the external stellar gas, but this generally would not happen before
the jet head has reached the outer layers of the progenitor star.  The length
of the cocoon region is similar to that of the jet, but its 
breadth is determined by its transverse velocity. 
A wide jet would create a near-spherical cavity, but a
narrow jet would advance much faster than the transverse expansion of the
cocoon (Matzner 2002), so that the cocoon would be ``cigar-shaped''
(or maybe more like an ``hourglass'' in the case when the external
pressure is a steep function of radius; see Fig. 1c).
  
At the radii $\approx r_*$ (probably the radius of the He core) where
the head of the jet starts to advance relativistically, the volume of
the material  deposited into the cocoon, 
$\Lambda_{\rm cav}$, is related to both jet and cocoon expansion velocities
by $\Lambda_{\rm cav} \approx (\pi/3) r_{*}^3 (V_c/V_h)^2$ (Matzner
2002). At that point in time, 
the cocoon plasma would itself be able to break out and
accelerate. Unlike the jet, this cocoon material does not have a
relativistic outward motion, although it has a relativistic internal
sound speed (i.e. similar energy to mass ratio). At first an asymmetric
bubble (since pressure balance may never be reached if the external
pressure falls off much steeper than $r^{-3}$) will be inflated which
can expand most rapidly along the rotation axis (Fig.~\ref{fig1}c) and
may eventually escape the stellar progenitor. But it may never expand
freely unless it escapes into an exponentially decreasing atmosphere
with $\beta \goa 5$ (.i.e. bare He star; see \S 3).

\subsection{Trapping and collimation}

Insofar as the cocoon material
and the lower-entropy stellar envelope can be treated as two separate
fluids (i.e. diffusion and viscosity can be neglected), it is
feasible to estimate the cone angle, $\theta_c$, of the expanding plasma.
The stellar envelope, which
contains the outflow along the axis, has 
a sufficiently large optical depth $\tau_{\rm env} \sim 10^{11}$ that
most of the radiation released is trapped, and transported into the
cavity by bulk flow rather than diffusing outwards. The optical
depth across the cavity is enormous, even if $\eta$ is so high that
the baryon density is low, because of the thermal pair
density ($T \ge$ 20 keV), and the radiation is then well enough trapped to
justify a fluid 
treatment. Collective plasma effects and magnetic fields may also
reduce the effective mean free path. Unless there is violent
entrainment, there would not be much mixing of baryons from the envelope
into the cocoon, so that  during the build-up its ratio of energy to
baryon-rest-mass will be given approximately by $\eta$.  If the
magnetic field contributes 
significant to the total energy density (i.e. MHD jet) and has a
preferred orientation, then the pressure and magnetosonic velocities
are of course anisotropic, but the dynamics would be essentially the
same as for pure radiation. When the pressure at the outer edges of
the cocoon cavity has halved the flow becomes transonic and the
cross-sectional area is minimised. In 
this way a directed nozzle can be established.  
The channel cross section is proportional to the total power
discharge and varies inversely with the pressure at the outer edges of
the cocoon cavity (Blandford \& Rees 1974). Unfortunately neither the
jet thrusts nor the pressures at the outer edges of the cocoon are
known well enough to
quantitatively predict the dimensions of the nozzle. Beyond the
nozzle, however, the external pressure drops steeply, and the cocoon
material expands freely in the transverse direction. The flow spreads
out over an angle $\sim \gamma_c^{-1}$. If its free expansion starts
just outside the nozzle, where the Lorentz factor of the cocoon
material $\gamma_c$ is only $\sim$ 2, then it
will spread over a wide angle and will develop into a roughly
semi-spherical blast wave.  The cocoon fireball
will then expand with $\gamma_c \propto r/r_{\rm cav}$, where 
$r_{\rm cav}\sim \theta_c r_*$ is the typical, initial
dimension that the fireball would have if it started out
spherically.

\section{The state of Cocoon Fireballs}

\subsection{A brief overview of the fireball model} 
Before turning to the question of how the kinetic energy of the cocoon
plasma is converted to radiation, it is worthwhile to summarise now
the essential features of the generic fireball scenario.

In the so-called standard fireball model (see 
Piran 1999 for a recent review),  it is conjectured that the fireball
wind, expelled by a 
central source of dimension $r_0 \approx 10^{5}-10^{6}$ cm (notice
than in the case of the cocoon fireball, the  relativistic
plasma is confined to a much extended cavity), accelerates at
small radii such that its Lorentz factor grows linearly with radius
until the entire fireball energy is converted into kinetic energy at
$\eta r_0$ (Cavallo \& Rees 1978; Goodman 1986;
\Pacz 1986; Shemi \& Piran 1990).  This energy must be converted to
radiation in an optically-thin region, as the observed bursts are
non-thermal.  The radius of transparency of the ejecta is 
\begin{equation}
r_{\tau}=\left({\sigma_T E_{4\pi} \over 4 \pi m_p c^2 \eta}\right)^{1/2},
\end{equation}
where $E_{4\pi}$ is the isotropic equivalent energy generated by the
central site. The inertia of the swept-up external matter decelerates the shell
ejecta significantly by the time it reaches the  
deceleration radius (M\'esz\'aros \& Rees 1997; Chevalier \& Li 1999): 
\begin{equation}
r_{\gamma}=\left( {(3-s) E_{4\pi} \over  m_{p}Ac^2 \eta^2}\right)^{1/(3-s)},
\end{equation}
where the external medium particle density is $n(r)=Ar^{-s}$, with
$s=0$ for a homogeneous medium $n(r)=n_{\rm ism}$, and $s=2$ for a wind
ejected by the stellar progenitor at a constant speed.
Given a certain external baryon density $n(r)$, the initial Lorentz
factor $\eta$ then strongly determines where both internal and 
external shocks develop (see e.g. Fig. 3 of  Ramirez-Ruiz, Merloni \& Rees
2001). Changes in $\eta$ will modify the
dynamics of the shock deceleration and the manifestations of the
afterglow emission.

\subsection{The propagation of cocoon fireballs}

The above summary describes the qualitative features of the generic 
fireball model. While similar scaling laws are also expected for the
evolution of the cocoon material, the physical conditions within and
around this plasma  are different, and so appreciable deviations from
the ``standard'' evolution are thus likely to occur. This variety of
propagation effects can substantially modify the emergent
radiation,  so that important constraints on the nature of the
source producing them  can be obtained simply by detecting these
changes.  For this reason, we now turn to consider the cocoon
properties in more detail.   

Consider a homogeneous fireball of
energy $E_c$, total mass $M_c$ initially confined to a cocoon cavity. 
Clearly, since $\tau > 1$, the initial fireball
will be an opaque sphere in thermal equilibrium, characterised by a
single temperature: $T_c \approx 400 g E_{c,52}^{1/4}r_{{\rm
cav},9}^{-3/4}$ keV in a spherical cocoon cavity, or  $T_c \approx 100
g E_{c,52}^{1/4} \Lambda_{{\rm cav},30}^{-1/4}$ keV 
in a ``hourglass'' cocoon. $g$=11/4 for $T_c
> m_ec^2$ (photons and pairs), and it drops to 1 when   $T_c \ll
m_ec^2$ (only photons). 

When the radiation energy dominates the evolution (i.e. $\eta \gg 1$), the
fluid expands under its own pressure such that its Lorentz factor grows
linearly with radius, $\gamma_c \propto r/r_{\rm cav}$. When the
fireball has a size of $r_{\eta_c}=\eta r_{\rm cav}$, all the internal
energy has been converted into bulk kinetic energy and the matter
coasts asymptotically with a constant Lorentz factor $\eta$. This is
perhaps true for a cocoon propagating into the exponentially
decreasing atmosphere of a carbon-oxygen or helium post-Wolf-Rayet
star and into the circumstellar environment beyond it, as in
curve  (a) of Figure 2. When there is a
remaining H envelope (with $\rho_H \propto r^{-\beta}$)
beyond the He core,  the deceleration radius of the cocoon fireball
may well be inside the star. The energy required to sweep up an
external stellar mass of $m_{\rm env}$ is $\gamma_c^2 m_{\rm env}c^2$, where
$\gamma_c \propto r$ and $m_{\rm env} \propto \rho(r)r^3 \propto
r^{-\beta + 3}$. Thus, the cocoon blast wave is undecelerated for
$\beta \goa 
5$. By the same token, the initial relativistic jet will expand freely
provided  $\beta \goa 3$. For a moderately dense H envelope, the cocoon 
fireball, which starts being  decelerated by the stellar matter at $r
<r_\eta$ (see curve (b) of Fig. 2),
emerges from the H envelope with a Lorentz factor $\gamma_c < \eta$. 
For an extended or denser H envelope, corresponding to curve (c) in
Figure 2,  the 
cocoon of relativistic material would be stalled before
emerging. However, even in this case, the cocoon may have more energy than
the binding energy of the envelope, and could give rise to a
``hypernova'' (Iwamoto et al. 1998; \Pacz 1998;
Wang \& Wheeler 1998). An 
accompanying GRB will be present depending on whether or not the jet is also
choked.

\subsection{The role of $\eta$}
The baryon load build-up in the cocoon reservoir, $M_c \approx
E_c/\eta c^2$, influences the fireball evolution by increasing the
opacity and thus delaying the escape of radiation. The $e^\pm \;$ pair
opacity, $\tau_p$,  decreases
exponentially with decreasing local temperature, and falls to
unity when $T_p \approx$ 20 keV. The
matter opacity, $\tau_b$, on the other hand, drops as
$r^{-2}$, and thus the escape temperature may drop far below $T_p$
if $\tau _b > \tau_p=1$ (Shemi \& Piran 1990; Piran 1999). Two
critical values for $\eta$ determine the order of these transitions: 
\begin{equation}
\eta_p = \left({3\sigma_T^2 E_c \sigma T_p^4 \over \theta_c^2 m_p^2 c^4 r_{\rm
cav}} \right)^{1/2} \approx 10^{10}\;E_{c,52}^{1/2}r_{{\rm
cav},9}^{-1/2}\theta_{c,0}^{-1}
\end{equation}   
and
\begin{equation}
\eta_b=\left({3\sigma_T E_c \over 2 \theta_c^2 m_p c^2 r_{\rm
cav}^2}\right)^{1/3} \approx 5 \times 10^{3}\; E_{c,52}^{1/3}r_{{\rm
cav},9}^{-2/3}\theta_{c,0}^{-2/3}.
\end{equation}

The effect of the baryons is only negligible when $\eta > \eta_p$ and
the evolution is that of a pure photon-lepton fireball ($\tau_p=1
>\tau_b$). If $\eta$ is less extreme,
there are two qualitative changes in the fireball's mode of
propagation.   First, 
the matter opacity becomes  important when 
$\eta_p > \eta > \eta_b \approx 5 \times 10^{3}\;E_{c,52}^{1/3}r_{{\rm
cav},9}^{-2/3}\theta_{c,0}^{-2/3}$. The comoving 
temperature, in this case, decreases far below $T_p$ before
$\tau=\tau_b$ reaches unity, yet the fireball continues to be
radiation dominated and most of the energy still escapes as radiation. 
Second, for $\eta$ smaller than $\eta_b$,
the fireball becomes matter dominated before it becomes optically
thin, and most of the initial energy its converted into bulk kinetic
energy (this is likely to 
be the common situation for the initial jet fireball; see \S 3.1). 
These two modes of propagation will be referred to in the
following as radiation dominated and matter dominated,
respectively.\\ 

\section{Massive progenitors, stalled cocoons and Fe lines} 

The properties of the stellar envelope determined in the previous
section have an important effect on the appearance of a jet
propagating through it. While the details remain uncertain,
preliminary calculations suggest that a relativistic jet can be launched
along the progenitor rotation axis (MacFadyen \& Woosley 1999; Aloy et
al. 2000; MacFadyen, Woosley \& Heger 2001). One would expect
baryon contamination to be the lowest near this axis, because angular
momentum flings material away from it and material with
low-angular momentum falls into the black hole (Fryer \& Heger 2000). 
The jet would be
expected to break free of the H envelope, and in principle lead to a
successful GRB, provided the central engine feeding time exceeds the
He-core crossing time and that the stellar pressure continues to drops
at least as fast as $p_H \propto
r^{-4}$. In addition to the stellar pressure required to stall the initial
jet, there is another, smaller pressure which can prevent the cocoon
material to expand freely. The cocoon fireball  will, in general, be easier to
confine than the initial jet, for
two reasons: first, the fraction of matter with a clear line of sight
along the rotation axis will be much reduced 
(i.e. $\Omega_c > \Omega_{\rm funnel} \goa \Omega_j $), and second, the
cocoon blast wave may never expand freely for an H envelope density
varying as $\rho_H \propto r^{-\beta}$ with $\beta < 5$.

The energy of the relativistic material that accumulated in the cocoon
while the jet was advancing subrelativistically is nevertheless much
larger than the binding energy of the H envelope. As
soon as the jet penetrates into the low-density envelope beyond $r_*$,
the cocoon plasma  would itself be able to ``break-out'' and expand
through the envelope along the direction of least resistance, which is
likely to be the rotation axis of the stellar progenitor (perhaps
further channelled as the jet penetrates further into the
envelope). The cocoon fireball, in this 
scenario, may start to decelerate  
before it becomes matter dominated. It starts being relativistic 
during the acceleration stage, but a self similar phase
could then begin after enough external material has been
collected. In the interim, the Lorentz factor will
drop faster than that of a cold fireball propagating
through a similar density profile would (Fig. 2). 
If the H envelope did not exist, then the cocoon fireball would reach the
outer edge of this envelope after a few seconds, corresponding to case
(a) in Fig. 2, where $t_{c,H} \approx r_{\rm cav}/c  +
r_H/(2c\eta^2)$. However, for very extended or slow rotating stars
(i.e. denser cores),  the cocoon  material would generally carry  less
energy and inertia than the stellar envelope; it 
would then take up to a few hours for it to reach the outer surface of the
star as it expands subrelativistically with a velocity of
the order of $V_c \approx c(E_c/M_{\rm
env}c^2)^{1/2} \approx  10^{9}\; (M_{\rm env}/M_\odot)^{-1/2}$ cm
s$^{-1}$ (the case of an ultimately choked cocoon).

As it expands relativistically (but notice that, as argued above, for very
extended or slow rotating stars the cocoon material may be stalled prior
to the acceleration of the fireball to relativistic velocities), 
the cocoon fluid cools with $T \propto (r/r_{\rm
cav})^{-1}$.  The coasting photons, whose local energy is $T$ are
blue shifted. An observer detects them with a temperature of $T_{\rm
obs} \propto \gamma_c(r) T(r)$.  Seeing that $T \propto r^{-1}$ and $\gamma_c
\propto r$, we find that during the acceleration stage $T_{\rm obs}
\approx T_c$, where  
\begin{equation} 
T_c \approx \left\{ 
\begin{array}{ll}
 70\; r_{*,11}^{-5/8}L_{j,50}^{1/8} ({M_{\rm
 env} \over M_\odot})^{-1/4}\;\;{\rm keV} & ({\rm spherical})\\ \\ 
  34\; r_{*,11}^{-1/2}L_{j,50}^{1/4} V_{c,9}^{-1/2}\;\; {\rm keV} & ({\rm hourglass}) \\  
\end{array} 
\right. 
\end{equation}
is the
initial black-body temperature of the opaque, cocoon cavity.
These photons would escape
freely after the expanding fireball becomes optically thin, which is
likely  to occur at some distance from the stellar surface and thus $T_{\rm
obs} \le T_c$. The escape temperature may drop far below $\gamma_c(r)T_p$ if
condition $\eta > \eta_b$ is not satisfied or if
the deceleration of the fireball takes place before the radius of
transparency (but it is always $\ge 4000 \gamma_c(r)$ K, the recombination
temperature).  A similar diagram to Fig. 2 can be drawn for $T_{\rm
obs}$ as a function of radius (see Fig. 3). 

In the case of a successful
break-through of the jet, a strongly decelerated, cocoon fireball could
result in a potentially interesting and observable phenomenon.  
Not only would a conventional ``long'' GRB be detectable, followed by a
standard afterglow, but also there would be, after some seconds or up
to a day, a secondary, almost thermal (see Goodman 1986) brightening ,
caused by the cocoon 
photospheric emission containing $\ge 10^{50}$ erg . The observed
duration of this emission would be of the order of
$r_{\tau_c}/ (c \gamma_c^2) \sim$ a few hours for a cocoon
expanding at mildly relativistic velocities (case (b) in Fig. 3). 
A choked cocoon will, however, expand more or less isotropically
through the rest of the 
envelope, causing its disruption. This would then appear, after the
disrupted envelope becomes optically thin, as a type II supernova
(or a type Ib/c if there is a significant injection of radioactive
material). Some evidence for supernova-type emission has been found in: GRB
980326 (Bloom et al. 1999);  GRB 970228 (Reichart 1999); GRB 990712
(Bj\"ornsson et al. 2001);  GRB 000911 (Lazzati et 
al. 2001); GRB 011121 (Bloom et al. 2002; Dado et al. 2002; Garnavich
et al. 2002).  The 
decelerating external shock of a cocoon fireball may produce a
significant absorption edge in the cooled shocked ejecta, so long as
cooling is faster than adiabatic losses and protons are well coupled
to the electrons. Absorption edges can also arise from cooler, denser
material or filaments  in pressure equilibrium with the shocked
envelope ejecta (\Mesz \& Rees 1998).

It is, of course possible that the initial jet is a
magnetically confined  configuration, whose collimation properties are
unaffected by the distribution of the external matter. 
In this case, the cocoon cavity
would still have a dynamically-important magnetic field strength.
If the field were tangled, continuing reconnection processes may lead to
acceleration of non-thermal electrons. It is therefore plausible
that a substantial fraction of the energy stored in the cocoon cavity
could be released, via magnetic dissipation,  in a non-thermal $\sim$
UV/X-ray continuum with $L \approx E_c/t_{c,H} \sim 10^{47}$ erg
s$^{-1}$.  A magnetic field of $10^5$ G could confine clumps of gas
with densities up to $n \ge 10^{17} {\rm cm}^{-3}$, even at keV
temperatures. Such clumps would be
optically thick and could reprocess a non thermal UV/X-ray continuum
arising from within the dilute plasma between them (as envisaged by \Mesz \&
Rees 2001) and (in the absence of both entrainment and substantial
$e^{\pm}$ pair production) may also excite Fe-line emission. If the UV/X-ray
continuum can maintain a ionization parameter $\xi
=L_x/(nr^{2})$ of the order of $10^{3}-10^{4}$, a modestly supersolar
Fe mass fraction could yield a recombination line luminosity
comparable to the one observed in GRB 991216 (Piro et al. 2000; Ballantyne
\& Ramirez-Ruiz 2001; Vietri et al. 2001; McLaughlin et al. 2002;
Ballantyne et al. 2002; Kallman, M\'esz\'aros \& Rees 2002). 
 
\section{Compact progenitors, GRB precursors, and afterglow signatures}

A relativistic cocoon fireball  is likely to escape if 
the star loses its hydrogen envelope before collapsing (MacFadyen \&
Woosley 1999; Aloy et al. 2000; Wheeler et al. 2000; MacFadyen et
al. 2001; Matzner 2002). This is expected for 
example in stars with high radiative mass-loss (e.g. Ramirez-Ruiz et
al. 2001). The fireball cocoon will then emerge from the He core into
an exponentially decreasing atmosphere and into the rarified circumstellar
environment beyond it, where it acquires the limiting bulk Lorentz
factor $\gamma_c \approx \eta$, unless of course, $\eta > \eta_b$. In
this later case, the cocoon plasma  continues to be
radiation dominated  when it becomes optical thin. At this stage, the
baryons will switch immediately to a coasting phase with $\gamma_c
\approx r_{\tau_c}/r_{\rm cav}$ (where $r_{\tau_c}< r_\eta$) and most of the
energy escapes as photons. As follows from  the previous discussion, 
an observer will detect them with a temperature of  $T_{\rm obs} \propto
\gamma_c T$.  Thus the observed thermal peak frequency
would be in  the BATSE [20-600]~keV spectral window (see equation 7)
for compact He envelopes $r_* \approx 10^{11}$~cm (the outer edge of
the He core varies with initial mass, roughly as $r_* \sim 10^{11} (M_i/35
M_\odot)^{-2}$; Woosley, Langer \& Weaver 1993) .  For extended He
envelopes ($r_* \ge 10^{12}$~cm), however, the thermal emission could
be detectable with instruments like Ginga and the {\it Beppo}SAX wide
field cameras. The time delay between the cocoon photospheric emission
and the start of the main burst 
is given by $\Delta t_\gamma \approx  r_{\rm cav}/(2c) +
r_{\tau_c}/(2c\eta^2)  - \max[{r_{\tau_j}},r_{\rm int}]/(2c\eta^2)$, where
$r_{\rm int}$ is the radius of internal shocks in the jet fireball. The thermal
signal emerging from a cocoon fireball with
$\eta_p > \eta > \eta_b$ will most likely appear as a transient signal 
at the beginning of the main burst since $|\Delta t_\gamma| \le 0.1$ s. 
In contrast to the main burst, this signal would only last for $r_{\rm
cav}/c \approx 0.1$ seconds. 
Such thermal precursors may have indeed been observed by Ginga (Murakami et
al. 1991). For compact stellar cocoons ($r_{\rm cav}<10^{10}$~cm),
however, this thermal signal could be very short lived
and may be difficult to disentangle from the internal shock emission.

The cocoon fireball will strongly modify the usual properties of the
standard internal shocks and the afterglow
emission when $\eta < \eta_b$ (i.e. the cocoon fireball becomes matter
dominated).  The total energy available in the cocoon  afterglow is
essentially, the kinetic energy of the relativistic  
wind deposited during the He-core traversal time, here estimated to be
 $\approx 7 r_{*,11}$ seconds, minus the fraction dissipated (and
converted into prompt $\gamma$-rays)   in internal shocks. The size of
the stellar cavity and the initial Lorentz factor $\eta$, along with the
jet lifetime, strongly  
determine whether or not the cocoon fireball would carry  less energy and 
inertia than the relativistic jet itself. For $E_{4\pi,c} >
E_{4\pi,j}$,  the main afterglow will be 
dominated by the deceleration of the cocoon fireball (this scenario is
likely to occur when the source lifetime is of the order of the
He-core traversal time, so that the energy carried by  the initial jet is much
less than that accumulated in the cocoon).  An interesting 
consequence of this scenario is that if the observer lies off-axis to
the jet, there could be a large fraction
of detectable afterglows for which no $\gamma$-ray event is detected
(commonly referred to as``burstless'' or ``orphan'' afterglows; see
M\'esz\'aros, Rees \& Wijers  1998). The initial jet starts being
decelerated by the external medium at a smaller radius, so that the
cocoon material always overtakes it and sweeps the 
jet material. This collision  would be an important
contribution to the observed afterglow at early times ($ \approx {r_{\gamma_j}
\over 2c \eta^2} < {r_{\gamma_c} \over 2c \eta^2}$), when the
afterglow emission produced by the jet fireball dominates. The
afterglow shock may experience,  after starting out in the canonical
manner, a ``resurgence" similar to the shock refreshment produced by the 
delayed energy injection of a long-lived central engine (see Panaitescu,
M\'esz\'aros \& Rees  1998 for the case of GRB 970508). 

If the jet
produced by the accretion maintains its energy for much longer than it
takes the jet to reach the surface of the H envelope, or is very
highly collimated, the initial jet
is likely to be more energetic than the cocoon material (i.e.  $E_{j} >
E_{c}$, and since entrainment may dominate in the initial fireball
$E_{4\pi,j}={4\pi E_{j} \over \Omega_j} > E_{4\pi,c}={4\pi E_{c} \over
\Omega_c}$).  Figure 4  shows the evolution of both jet and
cocoon fireballs for this matter dominated case. 
The adiabatic fireball evolution
(thick solid line) was computed using a similar numerical method to
the one developed by Kobayashi, Sari \& Piran (1999). 
The main afterglow will  be produced by the slowing down
of the jet as in the usual case, however, the emission at early stages would be
caused by the deceleration of the cocoon fireball. Figure 5 shows
the contribution of the cocoon afterglow radiation at low frequency
and at early times when
the initial jet contribution to 
the emission is negligible. If the external medium is homogeneous the
sub-millimiter afterglow produced by the jet should rise 
slowly at times between a few hours and one day, while for the cocoon
material the emission should fall steeply after this. Therefore
observations made at sub-millimiter frequencies with the SCUBA (James
Clerk Maxwell Telescope) or with MAMBO (IRAM telescope) instruments
would be very powerful in determining if the early afterglow is
dominated by the cocoon fireball emission.

There could also be additional precursor signatures which are not 
associated with the ejecta blast wave, but with the dynamics of the 
fireball at the coasting phase. The $\gamma$-ray (i.e internal shocks)
signal emerging 
from the cocoon fireball  would precede by $\Delta t_\Gamma \approx
{r_{\tau_j} \over 2c\eta^2} \approx$ a few seconds that produced by the
initial relativistic wind (provided that $E_{4\pi,j} >
E_{4\pi,c}$ and ${r_{\tau_j} \over 2c\eta^2} > r_{\rm cav}/c$).  
The observed variability time
scale of this prompt gamma-ray emission is related to the typical size
of the shocked plasma region containing the photon field:
$\Delta \approx r_{\rm cav}$ or $\Delta \approx r/\gamma^2$ for
$r > \Delta \gamma^2$. For compact stellar cocoons (i.e. $r_*<10^{11}$~cm),
the expected delay between the cocoon precursor
and the main pulse, $\Delta t_\Gamma$, would be proportional to the
total initial jet energy: $E_{4\pi,j}$. It will then simply reflect 
a correlation between the burst strength and the time elapsed
since the previous emission episode, similar to what is observed in
GRB lightcurves (Ramirez-Ruiz \& Merloni 2001). 

\section{Discussion}

We have discussed, in the context of a collapsar model of gamma-ray
bursts (which are normally assumed to occur in shocks taking effect
after the relativistic jet has broken free from the stellar envelope),
the dynamics and evolution of the relativistic plasma surrounding a
light, relativistic jet.  As the jet makes its way out of the stellar
envelope, most of its energy output during that period goes into a
cocoon of relativistic plasma. This material subsequently escapes
along the direction of least resistance. Provided that the density
along the H envelope vary monotonically with radius as $\rho_H \propto
r^{-\beta}$, the properties of the cocoon plasma would be similar to
those argued in \S 3. A collimated cocoon moving into a region
with $\beta \approx 5$ (i.e. no H envelope) will become overpressured
relative to its surroundings, and thereafter expands 
freely. If the relativistic jet 
carries less energy and inertia than the cocoon plasma itself (${4\pi
E_{c} \over \Omega_c} > {4\pi E_{j} \over \Omega_j}$), it will
start to decelerate at a smaller radius than the
collimated cocoon fireball, so that the latter would overtake it.
In this case, the afterglow would be dominated by the
emission of the cocoon material, which is likely to be ejected at larger angles
relative to the observer than those from the jet itself. 
On the other hand, if the jet
produced by the accretion maintains its energy for much longer than it
takes the jet head to reach the surface of the He envelope, the 
relativistic jet is likely to contain
substantially more energy than the off-axis cocoon material (since
$\Omega_j \ll \Omega_c$), so that 
it dominates the flux after expanding for a longer time than the
initially observed off-axis region.

Additional effects are expected
when the  cocoon fireball material becomes optically thin. Shock waves
within the plasma can contribute with a short-lived ($\approx$ few seconds)
non-thermal $\gamma$/X-ray transient  for a
cocoon propagating inside a compact post-Wolf-Rayet star (and so long
as $\eta < \eta_b \approx 5 \times 10^{3}\;E_{c,52}^{1/3}r_{{\rm
cav},9}^{-2/3}\theta_{c,0}^{-2/3}$; see \S 3). For very extended or slow
rotating stars,  a long lasting (few hours to a day)  UV/X-ray (almost) thermal
pulse, whose total  energy may be a few percent of the total 
burst energy, is likely to appear. If magnetic dissipation within
this plasma is important, it is also possible that a
substantial fraction of the energy stored in the cocoon can contribute  
a non-thermal UV/X-ray afterglow, and also excite Fe line emission from
the envelope gas. The detection 
of these prompt multi-wavelength signatures would be a test of the
collapsar model; and the precise measurement of the time delay between
emissions may help constrain the dimensions and properties of the H
envelope, the 
size of the cocoon cavity,  the initial dimensionless entropy of the
jet, $\eta$,  and the radius of the emitting region.   
The processes discussed here suggest 
that if GRBs are the outcome of the
collapse of massive stars involving a relativistic fireball jet,
bursts and afterglows may have a more complex spectra and 
time-structure than those alluded to the ``standard''  model.

\section*{Acknowledgements}
It is a pleasure to acknowledge many helpful  conversations with
Roger Blandford and Peter M\'esza\'ros. We thank the referee Chris Matzner
for numerous insightful comments and suggestions. 
This research has been supported by
CONACYT, SEP, ORS foundation, the Royal Society, and the Italian MUIR.

\bsp

\label{lastpage}

\newpage

\newpage

\begin{figure*}
\psfig{figure=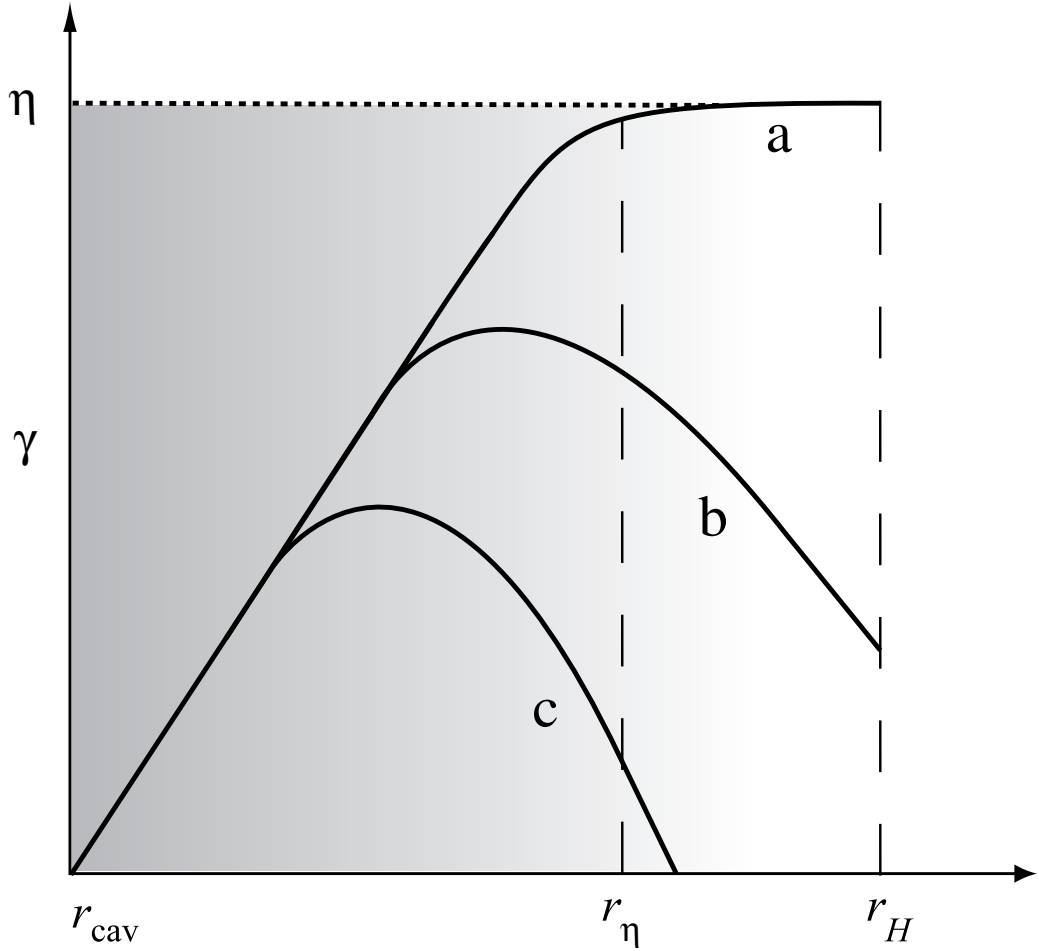,width=0.9\textwidth}
{\caption{
This diagram shows, for three illustrative cases, how the cocoon
expansion would be affected by the properties of the stellar envelope
through which it propagates. The axes (logarithmic) are $\gamma_c$
versus $r$,  where $r$ is
$r_{\rm cav} \approx r_*\theta_c$ when the cocoon  is observed to
start its expansion. Three illustrative cases are depicted. In case
(a), the stellar matter has a low density (i.e. exponentially
decreasing atmosphere of a carbon-oxygen or helium post-Wolf-Rayet
star), and the cocoon blast wave  sweeps up all the envelope's matter before
it has been decelerated. When the
fireball has a size of $r_{\eta_c}=\eta r_{\rm cav}$, all the internal
energy has been converted into bulk kinetic energy and the matter
coasts asymptotically with a constant Lorentz factor $\eta$.
In case (b), with higher stellar density, deceleration occurs at radii
$< r_{\eta_c}$, and the blast wave is still moving through the H envelope
material during the afterglow. After that, the cocoon escapes into the
circumstellar environment beyond the stellar envelope, where it
escapes freely with $\gamma_c < \eta$. For a very dense (or extended)
H envelope, corresponding to case (c), the cocoon  material would be
unable to break free from the stellar envelope. However, even in this
case, the cocoon material has more energy than the binding energy of
the envelope.} 
\label{fig2}}
\end{figure*}

\newpage

\begin{figure*}
\psfig{figure=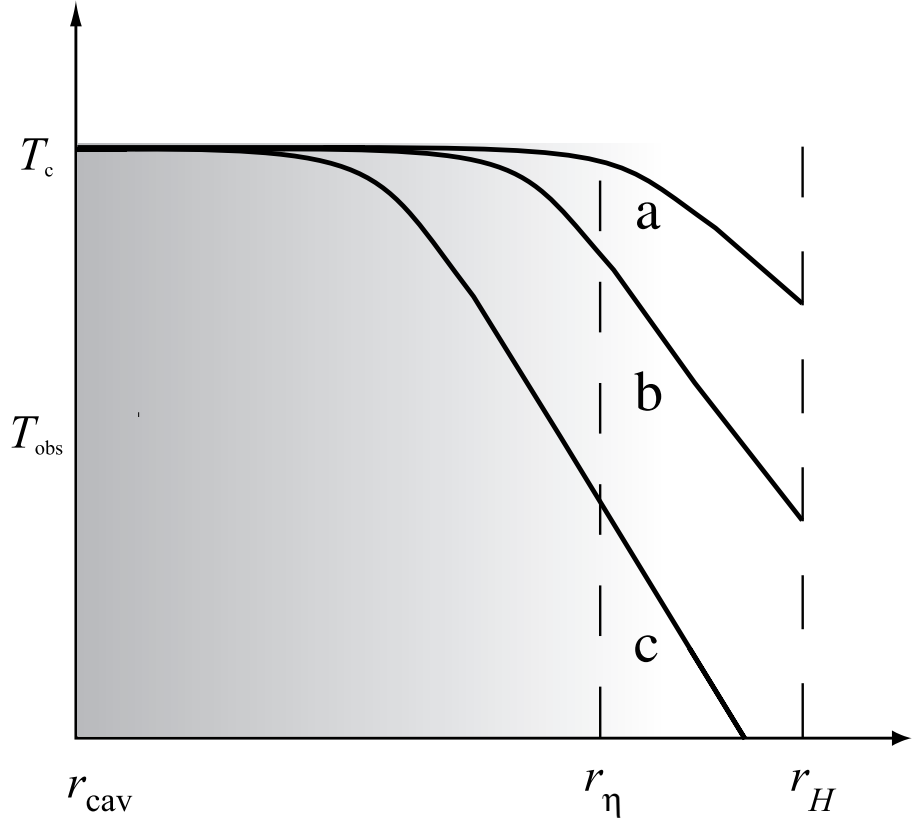,width=0.9\textwidth}
{\caption{
Schematic plot of the escape temperature $T_{\rm obs}$ as a function
of radius for the three, qualitatively different cases depicted in
Figure 2. As it expands, the cocoon fluid
cools with  $T \propto (r/r_{\rm cav})^{-1}$. The coasting photons,
whose local  energy is $T$ are  blue shifted. An observer would detect
them with a temperature of $T_{\rm obs} \propto \gamma_c(r)
T(r)$. These photons would escape freely after the expanding fireball
becomes optically thin, which is likely  to occur at some distance
from $r_H$.}
\label{fig3}}
\end{figure*}

\newpage
\begin{figure*}
\psfig{figure=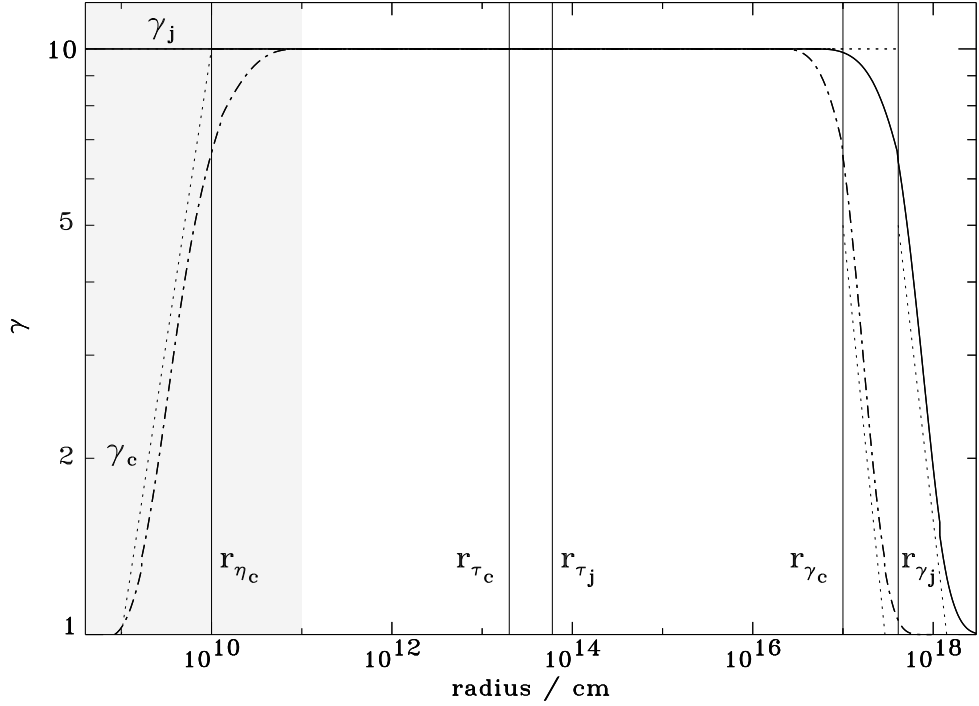,width=0.9\textwidth}
{\caption{Evolution of a matter dominated, cocoon fireball from its initial
formation at rest to its final stage. 
Both the initial jet (solid line) and the cocoon fireball (dashed
dotted line) propagate into the exponentially decreasing H atmosphere
and into the circumstellar environment beyond it, where they will both
accelerate while 
expanding and then coast freely until the surrounding matter will
eventually influence their coasting expansion. The energy dissipation is due to
interaction with the ISM via a relativistic forward shock and a
Newtonian reverse shock. The parameters for this
computation are: $\eta=10$, $r_{\rm cav}\sim 10^{9}$ cm,
$L_j=10^{50}$ erg s$^{-1}$, $n_{\rm ism}=1$ cm$^{-3}$, $\theta_j$=0.1
and $\theta_c$=1.0. Shown are the
numerical value of the average Lorentz factor (solid and dashed lines)  and its
analytical estimate (dotted line)}  
\label{fig4}}
\end{figure*}

\newpage
\begin{figure*}
\psfig{figure=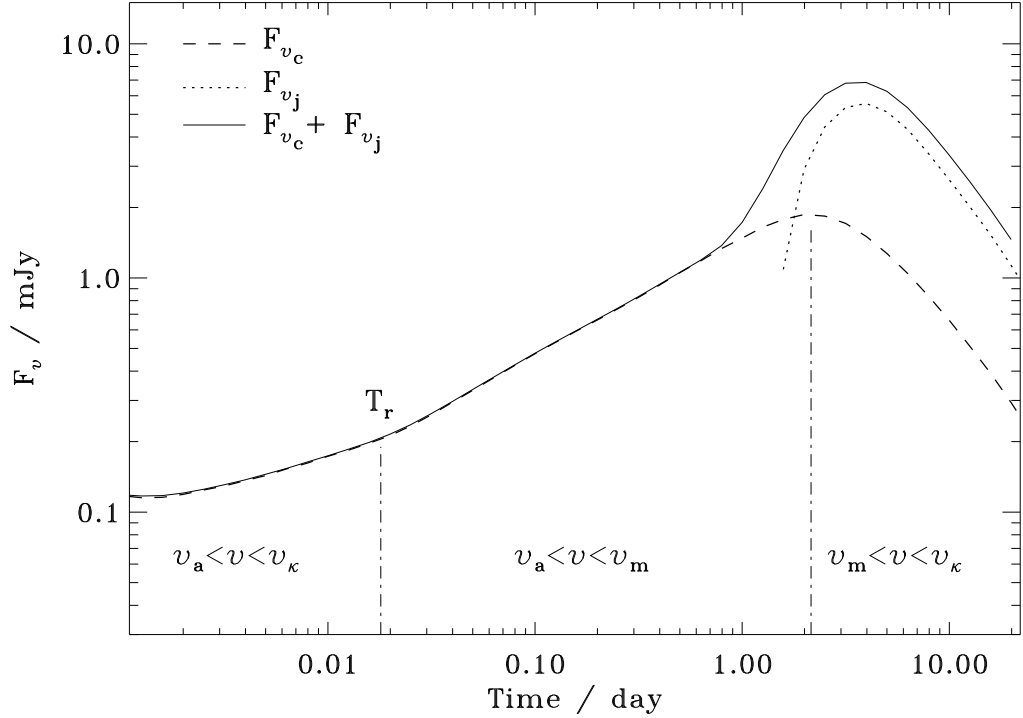,width=0.9\textwidth}
{\caption{Comparison between the initial jet and the cocoon afterglow
lightcurves  at early times and at observing frequency $\nu
=10^{12}$ Hz. The afterglow brightness depends on the
relationship between this frequency and those of the injection
($\nu_m$), cooling ($\nu_\kappa$), and absorption ($\nu_a$)
breaks (e.g. Panaitescu \& Kumar 2000). 
The dotted and dash lines represent the jet and cocoon
afterglow contributions  
corresponding to the parameters of the calculations
shown in Figure 4.}   
\label{fig5}}
\end{figure*}

\end{document}